# A QoS Provisioning Recurrent Neural Network based Call Admission Control for beyond 3G Networks

Ramesh Babu H.S.[1], Gowrishankar[2], Satyanarayana P.S[3].

[1]Department of Information Science and Engineering,
Acharya Institute of Technology
Bangalore, INDIA

[2]Department of Computer Science and Engineering,
B.M.S. College of Engineering,
Bangalore, INDIA

[3]Department of Electronics and Communication Engineering,
B.M.S. College of Engineering,
Bangalore, INDIA

**Abstract**

The Call admission control (CAC) is one of the Radio Resource Management (RRM) techniques that plays influential role in ensuring the desired Quality of Service (QoS) to the users and applications in next generation networks. This paper proposes a fuzzy neural approach for making the call admission control decision in multi class traffic based Next Generation Wireless Networks (NGWN). The proposed Fuzzy Neural call admission control (FNCAC) scheme is an integrated CAC module that combines the linguistic control capabilities of the fuzzy logic controller and the learning capabilities of the neural networks. The model is based on recurrent radial basis function networks which have better learning and adaptablity that can be used to develop intelligent system to handle the incoming traffic in an heterogeneous network environment. The simulation results are optimistic and indicates that the proposed FNCAC algorithm performs better than the other two methods and the call blocking probability is minimal when compared to other two methods.

**Keywords:** *Radio resource management, Heterogeneous wireless Networks, Call admission control, Call blocking probability, Recurrent radial basis function networks.*

## 1. Introduction

The majority researchers believe that the next stage beyond third-generation(3G) networks will include multiple wireless access technologies, all of which will coexist in a heterogeneous wireless access network environment[1,2] and use a common IP core to realize user-focused service delivery. The coexistence of Heterogeneous radio access technologies (RATs) will noticeably amplify the intensity different in development of different high-speed multimedia services, such as video on demand, mobile gaming, Web browsing, video streaming, voice over IP and e-commerce etc. Seamless inter system roaming across heterogeneous wireless access networks will be a major feature in the architecture of next generation wireless networks [3]. The future users of mobile communication look for always best connected (ABC) anywhere and anytime in the Complementary access technologies like Wireless Local Area Networks (WLAN), Worldwide Inter operability for Microwave Access (Wi-Max) and Universal Mobile Telecommunication Systems (UMTS) and which may coexist with the satellite networks [4- 6].It is very well evident that no single RAT can provide ubiquitous coverage and continuously high quality service. The mobile users may have to roam among various radio access technologies to keep the network connectivity active and to meet the applications/users requirements. With increase in offered services and access networks, efficient user roaming and management of available radio resources becomes decisive in providing the network stability and QoS provisioning.

The mobile communication networks are evolving into adaptable Internet protocol based networks that can handle multimedia applications. When multimedia data is supported by wireless networks, the networks should meet the quality of service requirements. One of the key challenges to be addressed in this prevailing scenario is the distribution of the available channel capacity among the multiple traffic ensuring the QoS requirements of the traffic that are operating with different bandwidth requirements.

There are many call admission control(CAC) algorithms proposed in the literature to handle single-class network traffic such as real-time traffic like voice calls [7-10].To serve the multiple classes of traffic we have the Partitioning CAC [11][12] and threshold based CAC [13] .The paper proposes the CAC framework for multi traffic based heterogeneous wireless networks . The resource allocation is a challenging task when the resources are always in scarce in a wireless environment. Efficient and intelligent call admission control policies should be in place which can take care of this contradicting environment to optimize the resource utilization. There are works reported on computation intelligence based call admission control algorithms. These algorithms admit or reject the call by applying computational intelligence techniques like fuzzy logic [14], Genetic algorithm [15], and fuzzy logic with Multi Attribute Decision Making (MADM) [16]. The combination of fuzzy and neural networks which forms a hybrid fuzzy neural network (FNN) is used for the radio resource management [17] .These intelligent techniques exhibit better efficiency which leads to higher user's satisfaction.

In this paper we propose a fuzzy neural approach based call admission control in a multi class traffic based Next Generation Wireless Networks (NGWN). The proposed FNCAC scheme is an integrated CAC module that combines the linguistic control capabilities of the fuzzy logic controller and the learning capabilities of the neural networks .The CAC model is developed using fuzzy Neural system based on Recurrent Radial Basis Function Networks (RRBFN) . RRBFN has better learning and adaptability that can be used to develop the intelligent system to handle the incoming traffic in the heterogeneous network environment. The proposed FNCAC can achieve reduced call blocking probability keeping the resource utilisation at an optimal level. In the proposed algorithm we have considered three classes of traffic having different QoS requirements and we have considered the heterogeneous network environment which can effectively handle these traffic. The traffic classes taken for the study are *Conversational traffic*, *Interactive traffic* and *Background traffic* which are with varied QoS parameters.

The further sections of the paper are organized as follows. The section II discusses on the soft computing techniques in RRM. Section III focuses on the Analytical model of the proposed call admission control based on higher order Markov chains. The section IV discusses the proposed intelligent FNCAC model. The section V represents the simulation results and conclusion is presented in section VI.

## 2. Soft Computing Techniques for RRM

The application of intelligent techniques has become wide spread for nonlinear time varying and complex problems that were posing a great challenge to researchers when they used the conventional methods. These soft computing techniques such as fuzzy logic, artificial neural networks and the hybrid systems like fuzzy neural networks have outperformed the conventional algorithmic methods. The advantages of these methods are many, which include most notably learning from experience, scalability, adaptability. Moreover, it has the ability to extract rules, without detailed or accurate mathematical modelling. All these features make the soft computing techniques the best candidates for solving the complex problems in any domain.

### 2.1 Fuzzy Logic

The concept of Fuzzy logic has been extensively applied in characterizing the behaviour of nonlinear systems. The nonlinear behaviour of the system can be effectively captured and represented by a set of Fuzzy rules [18]. Many engineering and scientific applications including time series are not only nonlinear but also non-stationary. Such applications cannot be represented by simple Fuzzy rules, because fixed number of rules can describe only time invariant systems and cannot take in to account the non-stationary behaviour. Recently, a new set of Fuzzy rules have been defined to predict the difference of consecutive values of non-stationary time series [19].

Advantages of Fuzzy Logic approach [20] are, easy to understand and build a predictor for any desired accuracy with a simple set of Fuzzy rules. Due to less computational demand there is no need of mathematical model for estimation and also for fast estimation of future values .The Limitations of Fuzzy Logic approach is, first it works on Single step prediction, second, the fuzzy logic do not have learning capability.

### 2.2. Neural Networks

The neural networks are low-level computational elements that exhibit good performance when they deal with sensory data. They can be applied to the situation where sufficient observation data is available. The Neural network methods are used in problems like control, prediction and classification. Neural Networks are able to gain this popularity because of the commanding capacity that they have in modelling exceptionally complex non linear functions. Neural networks have a biggest advantage in terms of easy to use which is based on training-prediction cycles. Training the neural networks plays a crucial role in the system usage of neural networks. The training pattern that contains a predefined set of inputs and expected outputs is used to train the neural networks.

Next, in prediction cycle, the outputs are supplied to the user based on the input values. To make the neural networks to behave like a physical system or predict or control, the training set used in the training cycle shall consist of enough information representing all the valid cases [21-23].

Neural Networks are flexible soft computing frameworks for modelling a broad range of nonlinear problems [24]. One significant advantage of the neural network based approach over other classes of nonlinear models is that NNs are universal approximation tools that can approximate large class of functions with a high degree of accuracy [25]. This approximation power of Neural Network model comes from several parallel processing elements, called as 'neurons'. No prior assumption of the model form is required in the model building process. Instead, the network model is largely determined by characteristics of the data. Single hidden layer feed forward network is the most widely used model for prediction and forecasting of time variant functions. The model is characterized by a network of three layers of simple processing unit connected by non-cyclic links. The architecture of feed-forward neural network is shown in Figure 1.

The relationship between the output $\hat{y}(t)$ and the inputs $\{y(t-1), y(t-2),...,y(t-n)\}$ can be mathematically expressed as [26],

$$\hat{y}(t) = w_0 + \sum_{j=1}^{Q} w_j g\left(w_{oj} + \sum_{i=1}^{n} w_{ij} y(t-i)\right) + e(t) \quad (1)$$

Where $w_{ij}(i=0,1,2,...,n, j=1,2,...,Q)$ and $w_j(j=0,1,2,...,Q)$ are model parameters often called connection weights, $n$ is the number of input nodes and $Q$ is the number of hidden nodes. $g(.)$ Represents a transfer function of the processing element, the transfer function can be logistic or Gaussian [26]. The NN model having a logistic or Gaussian transfer function can perform nonlinear functional mapping from the past observation to the future value $\hat{y}(t)$ i.e.

$$\hat{y}(t) = f(y(t-1),......y(t-n), W) + e(t) \quad (2)$$

Where $W$ is a vector of all input parameters and $f(.)$ is function determined by network structure and connection weights. Thus, the neural network model is equivalent to nonlinear auto regressive model.

The feed forward network can effectively model nonlinear time series. The time-varying wireless network parameters are represented as nonlinear and non-stationary time series. The recurrent connection in NN architecture is also called as 'short term memory' and will process the non-stationary behaviour of the time series.

The feed forward NNs can be divided into two classes: *static (non-recurrent)* and *dynamic (recurrent)*. In *Static NNs*, output is linear or nonlinear function of its inputs and generates the same output for a given input vector. These NNs are suitable for spatial pattern analysis. In this case, the relevant information is distributed throughout the spatial coordinates of the input vector. The spatial dependencies in the input data can be found in the areas of pattern recognition and functional approximation [27].

In contrast, dynamic NNs are capable of implementing memories which gives them the possibility of retaining information to be used later. The network can also generate diverse output in response with the same input vector, because the output may also depend on actual state of the memories. Dynamic NNs have inherent characteristic of memorizing the past information for long term or short term periods. These networks are ideal for processing spatio-temporal data.

The Recurrent Neural Network (RNN) architecture can be classified in to fully interconnected nets, partially connected nets and Locally Recurrent & Globally Feed-forward (LRGF) nets [28]. The fully connected networks do not have distinct input layer/nodes. Each node has input from all other nodes.

The partially connected RNN can be implemented by

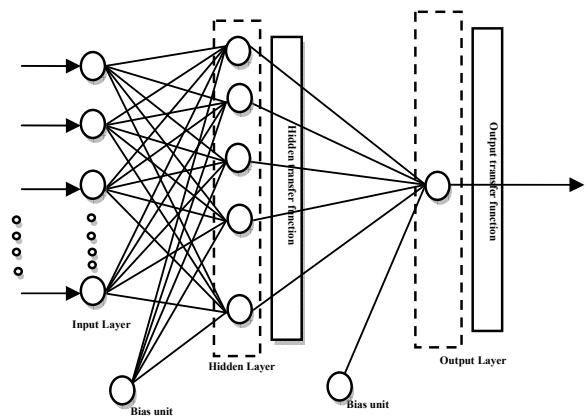

Figure 1  Feed forward neural networks

adding a feedback connection to the existing feed-forward NN to process the temporal information of the data. The feedback connection may be from hidden layer (Elman net) or from the output layer (Jordan net) [28, 29]. In the LRGF nets, self connecting neuron layer is either present in the input or on the output side of the feed-forward NN to process temporal information. The advantage of LRGF lies in its training algorithm. The standard gradient decent algorithm can be used to train the feed-forward NN for nonlinear functional approximations. A delayed input or the output from self connected neuron will  act as short term memory to process the time-varying information

There are good amount of work reported on the combination of neural and fuzzy logic approaches. This paper concentrates on the recurrent neural networks (RNNs) that have superior capabilities than the feed forward neural networks [31-32]. Since a recurrent neuron has an internal feedback loop to capture the dynamic response of a system without external feedback through delays, RNNs have the ability to deal with time-varying input or output through their own natural temporal operation [31]. Moreover; the RNNs are dynamic mapping and demonstrate good control performance in the presence of un modelled dynamics, parameter variations, and external disturbances [31-32]. The Radial Basis Function Network (RBFN) has a faster convergence property than a multilayer Perceptron (MLP) because the RBFN has a simple structure. Additionally, the RBFN has a similar feature to the fuzzy system. First, the output value is calculated using the weighted sum method. Then, the number of nodes in the hidden layer of the RBFN is the same as the number of 'if–then' rules in the fuzzy system. Finally, the receptive field functions of the RBFN are similar to the membership functions of the premise part in the fuzzy system. This makes the RBFN a very useful technique to be applied to control the dynamic systems. The implementation of RBFN bases using RBFN for recurrent RBFN based FNN improves the accuracy of the approximation function.

The benefits of neural network approach [24] are as follows. First, the NN Prediction accuracy is much superior to conventional approaches. Second, NN Model can be used for single and Multi step forecasting. Third, they are capable of learning the system and demands low computation structures. The limitations of NN approach are: The optimal choice of number of layers and number of neurons in each layer is by a heuristic process and it requires expertise in the field of NNs for a model designer. The deciding of the weights to the non-cyclic links will determine the accuracy of forecasting. Deciding the appropriate weights to the link is once again a heuristic process.

## 3. Analytical Model for CAC

In this paper we propose a novel analytical model admission control mechanism for reducing the call blocking probability there by increasing the resource utilization. This would achieve the Objective of guaranteeing the user QoS requirements. The proposed model is able to handle three types of the applications considered for the study which involves *conversation traffic, interactive traffic* and *background traffic*. All of these represent different QoS class of traffic with the required QoS parameters.

The Conversational traffic is sensitive to transfer delay and jitter. It demands guaranteed bit rate and low bit error rate. The examples of the applications belonging to this category are Video-conferencing and Audio conferencing. The Interactive traffic is a QoS Class that is not sensitive to Transfer Delay and Jitter but demands low Bit Error rate. The applications of this QoS class do not need Guaranteed Bit Rate, for example Web browsing, Interactive chats and Interactive Games. The Background traffic QoS class is not sensitive to transfer delay and jitter but needs low bit error rate from the network and these applications do not depend on guaranteed bit rate. The examples belonging to this group are E-mail, SMS applications. The assumption made for the design and development of analytical CAC model was type3 traffic which would require three channels to be assigned in the system and type2 traffic demands two channels and type1 traffic needs one channel.

The proposed model is developed keeping in mind the WCDMA, Wi-Fi, and Wi-Max .The CAC mechanism proposed is focused only on the system's ability to accommodate newly arriving users in terms of the total channel capacity which is needed for all terminals after the inclusion of the new user. In case when the channel load with the admission of a new call was precompiled (or computed online) and found to be higher than the capacity of the channel, the new call is rejected, if not, the new call could be admitted. The decision of admitting or rejecting a new call in the network will be made only based on the capacity needed to accommodate the call.

We consider a heterogeneous network which comprises a set of RATs $R_n$ with co-located cells in which radio resources are jointly managed. Cellular networks such as Wireless LAN and Wi-Max can have the same and fully overlapped coverage, which is technically feasible, and may also save installation cost. H is given as H {RAT 1, RAT 2, RAT k} where K is the total number of RATs in the heterogeneous cellular network. The heterogeneous cellular network supports n-classes of calls, and each RAT in set H is optimized to support certain classes of calls.

The Analytical model for CAC mechanism in heterogeneous wireless networks is modelled using Higher order Markov Model. In the proposed model it is assumed that, whenever a new user enters the network, it will originate the network request at the rate $\lambda_i$ and is assumed to follow a Poisson process. The service time of the different class of traffic and types of calls is $\mu_i$. The mean service time of all types of users were assumed to follow negative exponential distribution with the mean rate $1/\mu$. The Voice traffic is Erlang distributed, the condition that is considered for simulation is Negative Exponential distribution. The total number of virtual channels in the system are N. When the numbers of available channels are below the specified threshold the system will drop the calls. The threshold limit is determined by three positive integers $A_1, A_2$ and $A_3$.

These are called as *Utilization rates* where *A* is represented as $A = \frac{\lambda}{\mu}$.
Similarly,

$$A_1 = \frac{\lambda_1}{\mu_1}, \quad A_2 = \frac{\lambda_2}{\mu_2}, \quad A_3 = \frac{\lambda_3}{\mu_3}$$

are the *utilisation rate* of *type1* traffic, *type2* traffic and utilisation rate of *type3* traffic respectively. In general the values of the utilisation rate in a steady state system will be with in 1.

When the available number of channels falls below the threshold $A_3$ the proposed system will accept only the voice calls and web browsing. When the available number of channels falls below the threshold $A_2$ the proposed system will accept only the voice calls. If the available number of channels falls below the threshold $A_1$ the proposed system will not accept any calls as it reaches the stage where there will be no channels available to allocate to the incoming calls and leads to system blocking. The *P(0)* is the probability that there are no allocated channels in the designated system. The parameters of analytical performance model are also called as Performance model parameters. The parameters are number of *virtual channels (N), user arrival rate (λ), arrival rate of type 1 call (λ1), arrival rate of type 2 call (λ2.) arrival rate of type 3 call (λ3)* and *service time* of the calls is taken as $\mu_1$, $\mu_2$ and $\mu_3$.

Assuming that the arrival time of all the types of traffic are equal i.e. $\lambda_1 = \lambda_2 = \lambda_3 = \lambda$ and the service time for the types of traffic are equal i.e. $\mu_1 = \mu_2 = \mu_3 = \mu$, the call blocking probability for type1 traffic could be expressed as

$$P_n = \frac{a}{3}(P_{n-1} + P_{n-2} + P_{n-3}) \quad (3)$$

Where a = λ / μ which should be generally less than one for the system stability. Similarly, the call blocking probability for type2 traffic $P_{n-1}$ is

$$P_{n-1} = \frac{a}{3}(P_{n-2} + P_{n-3} + P_{n-4}) \quad (4)$$

And the call blocking probability for type3 traffic $P_{n-2}$ is represented as

$$P_{n-2} = \frac{a}{3}(P_{n-3} + P_{n-4} + P_{n-5}) \quad (5)$$

The call blocking probability for the overall system traffic $P_{nb}$ can be expressed as

$$P_{nb} = \frac{a}{3}(P_n + P_{n-1} + P_{n-2}) \quad (6)$$

## 4. Fuzzy Neural Call Admission Controller (FNCAC)

Our proposal to deal with the complex problem of call admission control in heterogeneous wireless network environment supporting multimedia traffic is developed using the hybrid model. This hybrid model is evolved by combining the fuzzy logic which is easy to understand and uses simple linguistic terms and if-then rules with the neural networks which are smart enough to learn the system characteristics. Therefore the Fuzzy neural networks combine the benefit of both neural networks and the fuzzy systems to solve the CAC problem. This research work particularly uses the feed forward neural networks which has the ability to map any nonlinear and non-stationary function to an arbitrary degree of accuracy [24].One such popular feed-forward network is the Radial Basis Function Network(RBFN). It is a single hidden layer feed-forward network. Each node in the hidden layer has a parameter vector called as *centre*. These *centres* are used to compare with network input and produce radically symmetrical response. These responses are scaled by connection weights of the output layer and then produce network output, where Gaussian basis function is used and given by equation (7).

$$\hat{y} = \sum_{i=1}^{n} w_i \exp\left(-\frac{\|y - \mu_i\|}{2\sigma_i}\right) \quad (7)$$

Radial Basis Function (RBF) has achieved considerable success in nonlinear function prediction but the performance of RBF is less satisfactory for the nonlinear and non-stationary function prediction[27].Recurrent Radial Basis Function Network(RRBFN) is a class of locally recurrent & globally feed-forward (LRGF) RNN. In LRGF network the recurrent/self-connection is either in the input layer or in the output layer. RRBFN is having recurrent connection at the input layer. Where $\sigma_i$ is the dimension of the influence field of hidden layer neuron, $y$ and $\mu_i$ are input and prototype vector respectively. The Recurrent Radial basis function network considers the time as an internal representation and the non-stationary aspect of nonlinear function can be obtained by having self-connection on the input neuron of sigmoidal firing function .The recurrent weights are in the range [-1 , +1 ] with normal distribution. This is a special case of locally recurrent, globally feed-forward neural network [28]. The RRBFN output for Gaussian basis function is as indicated in (8).Where $\hat{y}(.)$ is the predicted time series, n is the number of step prediction and j is the number of neurons in the input layer of RRBFN system

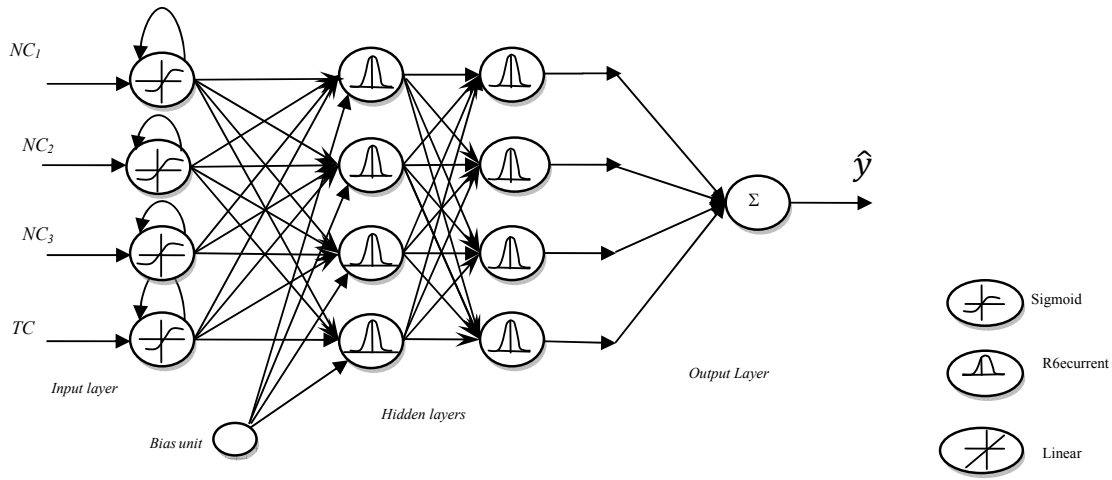

Figure 3  Fuzzy Neural CAC (FNCAC) model

$$\hat{y}(n) = \sum_{i=1}^{n} w_i \exp\left(-\frac{\sum_{j=1}^{m}(y^j - \mu_i^j)^2}{\sigma_i}\right) \quad (8)$$

The proposed architecture of RRBFN based FNCAC model is shown in Figure 3. The FNCAC takes the network characteristics of the three networks taken for the study and the requirements of the incoming traffic is taken as inputs. The cost is considered as the bias input. The neural network based Call admission control involves training and testing of RRBFN based CAC controller. The training and testing samples are randomly picked from the sample size of 1000. The RRBFN network has four layers: input, two hidden layers and output layer. For the training and testing, we have used 250 neurons in the input layer with sigmoid activation function and with the recurrent connections. The range of recurrent weights is -1 to +1. The hidden RRBF layers have 200 neurons with RBF activation function and output layer has single neuron with linear activation.

## 5. Simulation Results and Discussion

In this section, we present the numerical results and compare the call blocking probabilities of the different types of traffic. The set of experiments were conducted with varying the aggregate traffic and individual traffic of the network and the call blocking probability of Fuzzy neural technique was compared with the conventional CAC and Fuzzy based CAC. The aggregate utilization rate of the calls was considered with the call blocking probability of the FNCAC, conventional CAC and Fuzzy based CAC. As the combined traffic intensity increases the utilization rate also increases. The Fuzzy neural CAC model exhibits better performance in reducing the call blocking probability of the aggregate traffic which is assumed to have the varied traffic component of *type1*, *type2* and *type3* traffic. The performance comparison of fuzzy neural method, convention CAC and fuzzy based CAC is plotted in figure 4.

The next set of experiments was conducted to compare the call blocking probabilities of the individual traffic in Fuzzy neural based CAC. The type1 traffic has minimal call blocking probability when compared *type2* and *type3* traffic and also *type3* traffic has higher call blocking probability when compared to *type2* and *type1* traffic. The simulation results in figure 5 shows that the call blocking probability of the individual types of traffic will increase with the increase in the utilisation rate. The next set of experiments were conducted by considering only one type of traffic and the call blocking probability of the system was plotted for Fuzzy neural technique in comparison with the conventional CAC and Fuzzy based CAC. The graph in figure 6 considers only *type1* traffic in the system, figure 7 indicates the blocking probability of *type2* traffic for all the three systems. *Type3* traffic is considered independently in the system and call blocking probability was studied and is represented in figure 8. The study clearly indicates that the performance of the FNCAC is better than the other two CAC methods in terms of reduced call blocking probability.

## 6. Conclusion

In this paper, the performance of FNCAC system for next generation networks is compared and validated with the performance of fuzzy based CAC and conventional CAC. The Performance of FNCAC model in the heterogeneous RATs supporting multimedia traffic is studied pitching upon the call blocking probability by varying the utilization rate of the aggregate traffic and the individual traffic. The simulation study conducted records the following observations. The increase in the utilisation rate increases the call blocking probability of the system for both the aggregate traffic and the individual traffic. The experiment results indicate that the fuzzy neural CAC reduces the blocking probability by around 20% less compared to other two methods.

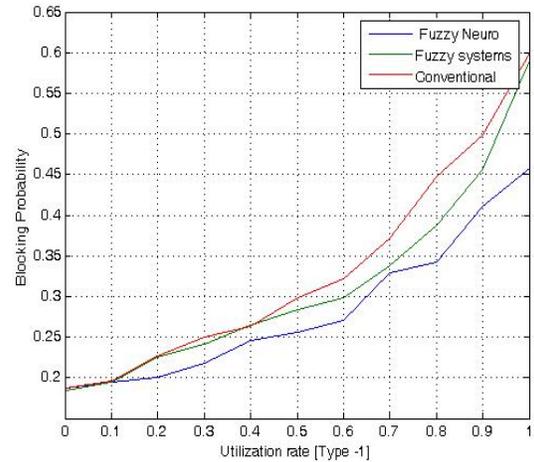

Figure 6 Call blocking probabilityforType1 Traffic

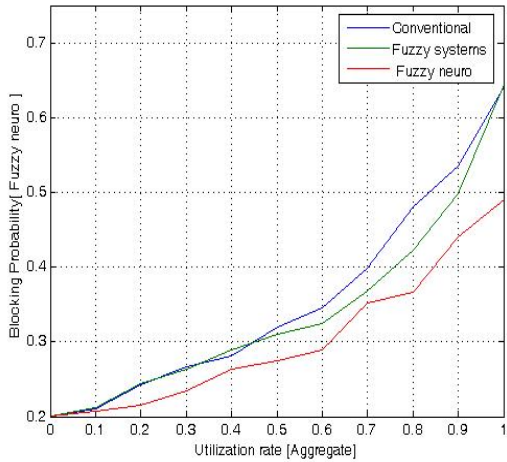

Figure 4 Call blocking probability for the Utilization rate (aggregate)

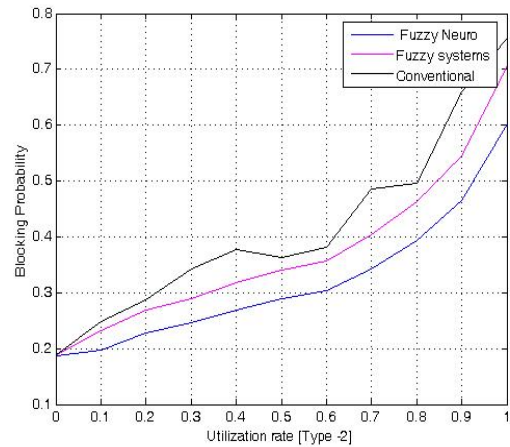

Figure 7 Call blocking probability for type 2 traffic

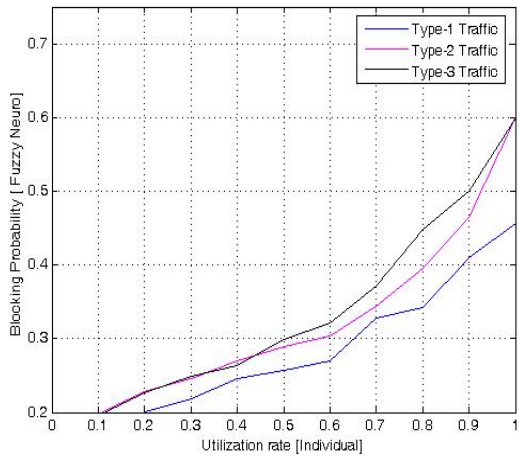

Figure 5 Individual traffic utilisation rate v/s FNCAC call blocking probability

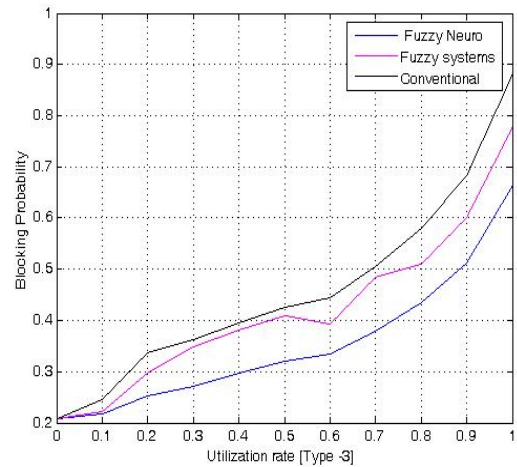

Figure 8 Call blocking probability for type3 traffic

## AUTHORS

**Ramesh Babu.H**.**S** received Bachelor of Engineering degree in Computer Science and Engineering from Bangalore University and MS in Software Systems from BITS, Pilani and pursuing Doctoral degree in Visvesvaraya Technological University, Belgaum. He is currently working with the Department of Information Science and Engineering, Acharya Institute of Technology, Visvesvaraya Technological University, Soladevanahalli, Bangalore-560 090, Karnataka, INDIA.

**Dr.Gowrishankar** received Bachelor of Engineering degree in Computer Science and Engineering from Mysore University and ME in computer Science and Engineering from Visvesvaraya Technological University and PhD from Visvesvaraya Technological University Belgaum. He is currently working with Department of Computer Science and Engineering, B.M.S. College of Engineering, Visvesvaraya Technological University, Bull Temple Road, Bangalore-560 019, Karnataka, INDIA.

**Dr.P.S.Satyanarayana** was Professor in Electronics and communication Engineering Department, B.M.S. College of Engineering, Visvesvaraya Technological University, P.O. Box. 1908, Bull Temple Road, Bangalore-560 019, Karnataka, INDIA.